\documentclass[apjl]{emulateapj}
\usepackage{amsmath,amsfonts,amssymb}
\usepackage[usenames]{color}
\definecolor{violet}{rgb}{0.05,0.0,0.4}
\usepackage[%
pdftitle={super structures},%
pdfauthor={Hu Zhan},%
bookmarksnumbered=true,%
linkcolor=blue,citecolor=violet,urlcolor=blue,colorlinks=true]{hyperref}
\allowdisplaybreaks[3]

\def\beq{\begin{equation}}
\def\eeq{\end{equation}}
\def\muK{\,\mu\mathrm{K}}
\def\wmap{\emph{WMAP}}

\def\lcdm{$\Lambda$CDM}
\def\deg{^\circ}
\def\Mpc{\,\mathrm{Mpc}}
\def\Mpch{\,h^{-1}\mathrm{Mpc}}

\def\RS{Rees--Sciama}
\def\SZ{Sunyaev--Zel'dovich}
\def\SLt{Sandage--Loeb test}

\def\rI{r_{\mathrm{I}}}
\def\rII{r_{\mathrm{II}}}

\def\rhoEi{\rho_{\mathrm{EdS},i}}

\def\rhoIi{\rho_{\mathrm{I},i}}
\def\dI{\delta_\mathrm{I}}
\def\dIi{\delta_{\mathrm{I},i}}
\def\KI{K_\mathrm{I}}
\def\rmo{\mathrm{o}}
\def\rme{\mathrm{e}}

\newcommand{\eqa}[1]{\begin{align} #1 \end{align}}
\newcommand{\reffig}[1]{Figure~\ref{#1}}
\newcommand{\refeq}[1]{Equation~(\ref{#1})}
\newcommand{\refsec}[1]{Section~\ref{#1}}
\newcommand{\refapp}[1]{Appendix~\ref{#1}}

\newcommand{\remove}[1]{}

\begin{document}

\title{
Rees--Sciama effect and impact of foreground structures on galaxy redshifts}
\shorttitle{\RS{} effect and galaxy redshifts}
\author{Hu Zhan} 
\shortauthors{Zhan}
\affil{Key Laboratory of Optical Astronomy, 
National Astronomical Observatories, Chinese Academy of Sciences,
Beijing 100012, China; zhanhu@nao.cas.cn}

\begin{abstract}
We estimate the \RS{} (RS) effect of super structures on the cosmic 
microwave background (CMB) temperature fluctuations and identify a 
related effect on galaxy redshifts. By numerically solving the geodesic 
equation, we find 
that both superclusters and supervoids can decrease the temperature of 
the CMB by several micro Kelvin in the central region and increase the 
temperature slightly in the surrounding area due to the RS effect. 
The two components of the RS effect, redshift and 
gravitational time delay, largely cancel each other, leaving an 
equivalent but much smaller effect on the CMB photons that started out at 
the same time from the distorted last scattering surface. For galaxies, 
the time delay effect is separable from the redshift effect, 
and the slight change to the redshift induced by super structures can 
be at the percent level of large-scale rms bulk velocities, which might 
only be detected statistically. 
On much smaller scales, a tiny redshift difference between two images 
of a strongly lensed source should exist in general, which 
is related to the Hubble expansion rate at the source redshift. 
However, as \citet{loeb98} pointed out, 
observational issues and the proper motion of the structure would make 
such a measurement impossible.
\end{abstract}


\section{Introduction}\label{sec:intr}
Measurements of the cosmic microwave background (CMB) with the
Wilkinson Microwave Anisotropy Probe \citep[\wmap,][]{bennett03}
have played a crucial role in establishing the concordance cosmological
model, \lcdm{} -- a flat cold-dark-matter universe with roughly 
three quarters of its content in the form of dark energy 
\citep[e.g.,][]{spergel03}. Besides the primary anisotropies, one also 
expects secondary CMB anisotropies arising from late-time gravitational 
effects, such as the integrated Sachs--Wolfe (ISW) effect 
\citep{sachs67}, and scattering processes, such 
as the thermal Sunyaev--Zel'dovich effect \citep{sunyaev72}.
Detections of these effects \citep*[e.g.,][]{birkinshaw91,
carlstrom96, fosalba03, boughn04,afshordi04, padmanabhan05, 
giannantonio08, granett09} 
are useful confirmations of the \lcdm{} model. 

The ISW effect on CMB photons is caused by the time variation of 
gravitational potentials as the photons travel through them. In 
order to have a measurable ISW effect, the photon travel time 
must be an appreciable fraction of the time scale, over which the 
potentials vary significantly. On large scales, overdensities in 
an Einstein--de Sitter (EdS) universe grows at the same rate as the
cosmic expansion, so that the linear ISW effect vanishes. However, 
in a \lcdm{} universe, the cosmic expansion rate exceeds the linear 
growth rate, causing a decay of potentials. A photon traveling 
through a decaying potential well (wall) gains (loses) energy.
Therefore, detection of the linear ISW effect provides a piece 
of evidence for dark energy in a flat universe. 

Nonlinear clustering causes extra evolution of the potentials 
against the background expansion, so that the ISW effect is 
present even in an EdS universe. This effect is also known as 
the \RS{} (RS) effect \citep{rees68}. On large scales,
the nonlinear contribution (or the intrinsic RS effect) 
to the full ISW effect is expected to be subdominant 
to the linear ISW effect in a cosmological constant (or dark 
energy) dominated universe \citep*{seljak96,tuluie96}. 

In a recent paper, \citet*{granett08} report a mean CMB 
temperature change of $|\Delta T| = 9.6\pm 2.2 \muK$ caused 
by superclusters and supervoids at $z \sim 0.5$ (hereafter,
we drop the prefix ``super'' where there is no ambiguity). 
These structures have radii of $\sim 100\Mpch$
or $4\deg$ at $z \sim 0.5$. The scales involved are large enough, 
so that the linear ISW effect should be the major source of the 
temperature change in the \lcdm{} model. 
However, the result is somewhat puzzling, because the temperature 
change is twice of what is estimated from simulations 
\citep{granett08}. Furthermore, there is no significant 
difference between the mean temperature of the clusters and 
that of voids in the reconstructed ISW map \citep*{granett09}. 

Given the studies above, it is of interest to reexamine the 
different mechanisms, through which cosmic structures alter the 
CMB temperature. We focus on the intrinsic RS effect in this 
paper. Our goal is not to reconcile the reported mean CMB 
temperature change with theory, but we note that the thermal \SZ{} 
effect and the proper motion effect of structures 
\citep{birkinshaw83,gurvits86,stebbins88} are not likely to explain 
the discrepancy. In \wmap{} bands, unresolved thermal 
\SZ{} signals would reduce the CMB temperature of clusters 
with respect to that of voids and hence suppress 
$|\Delta T|/T$, whereas the proper motion of a cluster or void 
would produce a dipole pattern without 
affecting the mean temperature.

We adopt the EdS universe as the background cosmology
to isolate the intrinsic RS effect (we drop
the word ``intrinsic'' hereafter). 
Because the growth rate of structures varies with the assumed
cosmology, the RS effect is model dependent. For
example, the combined RS and proper motion effects decrease
by a factor of $\sim 2$ from the EdS universe to an open  
universe with matter fraction $0.3$ \citep{tuluie96}. 

One approach to investigate the RS effect is to analyze 
its power spectrum \citep[e.g.,][]{seljak96}. 
The linear ISW effect increases the CMB 
temperature fluctuation and hence the power of the modes by 
raising the temperature of overdense regions and lowering that of 
underdense regions. Because the RS effect 
produces a net temperature decrement for both clusters and voids
(see \refsec{sec:gal} for an explanation), 
it would not increase the temperature fluctuation as much even if 
it had the same amplitude as the linear ISW effect. Thus, power
spectrum analyses may underestimate the importance of the RS 
effect. 

Ray-tracing through $N$-body simulations can 
provide a realistic estimate of the full ISW effect 
\citep[e.g.,][]{tuluie96, maturi07, cai09, cai10, smith09}
To do so exactly, one needs to solve for 
the metric and its spatial-temporal derivatives, 
which appear in the geodesic equation. The ISW effect is 
roughly $|\Delta T|/T  = |\Delta z|/(1+z) \sim 10^{-6}$--$10^{-5}$ 
for super structures, so that a redshift error of the order 
$10^{-7}$ or larger accumulated along the photon geodesic, 
or a comparable error from approximating the photon geodesic 
with a straight coordinate ray, can be a significant contamination 
to the results. Therefore, it is computationally challenging to 
trace photon geodesics in $N$-body simulations.

In this paper, we study the effect of individual structures 
with toy models \citep[see, e.g.,][]{thompson87, martinez90}.
The advantage of this approach is that (1) for certain class of 
models, one can obtain analytic solutions of the underlying metric 
without worrying about spatial, temporal, or mass resolutions
and (2) in models 
with symmetry, certain null geodesics can be computed in different 
ways, so that one can examine the precision of the calculations.
Specifically, we model the clusters and voids with the 
spherically symmetric Lema{\^i}tre--Tolman--Bondi 
\citep[LTB,][referenced herein]{lemaitre33,
tolman34,bondi47} solution and solve generic photon geodesics through
these structures 
\citep[see also][]{panek92,alnes06b,marra07,valkenburg09}.

Separately, the RS effect has been studied for compensated voids
and clusters \citep{inoue06,tomita08,sakai08} and for compensated 
shells \citep*{afshordi11} in the \lcdm{} universe with thin-shell 
approximation, perturbative calculations, and numerical calculations. 
The focus of this paper is on the effect caused solely by the 
evolution of the structures, so we do not include the cosmological 
constant $\Lambda$ except for a simple case in \refsec{sec:con}.
Our results are in qualitative agreement with those in the 
aforementioned works, and we extend the  
study to uncompensated structures as well.

As \citet{rees68} pointed out, redshift and time delay are the two
major components of the RS effect. The redshift component includes 
both compensation for the time delay component and the evolution of 
the potential. Since the effect of the potential evolution is 
usually subdominant, the two components of the RS effect often
shift the CMB temperature with comparable magnitudes
but in opposite directions. 

Galaxy redshifts are directly measurable, so fractional changes to 
galaxy redshifts due to intervening structures are larger
than that to the CMB temperature. This foreground-induced change of 
galaxy redshifts is closely related to but different from the 
RS effect or the ISW effect in general because of the separation of 
the redshift effect and the time delay effect on galaxies. 
However, unlike 
the CMB, which has a standard temperature, galaxies are spread over 
redshift space. Detecting such a redshift change is far more 
difficult than that of the RS effect. 

Images of a strongly lensed source should have slightly different
redshifts in general. Despite the extreme difficulty in measuring
such redshift differences \citep{loeb98}, 
we find in the ideal case that the redshift 
difference between two images of the source is considerably 
larger than that between two epochs of observations of the same image
separated by the amount of the time delay between the two images.
The latter is known as the \SLt{} \citep{sandage62,loeb98}.

The rest of the paper is organized as follows. \refsec{sec:ltb}
provides a brief introduction to the LTB solution and describes 
models of clusters and voids. \refsec{sec:geo} gives
the details of solving the null geodesics numerically. Results of
the RS effect on CMB temperature profiles and tiny perturbations 
to galaxy redshifts caused by intervening structures 
are presented in Sections \ref{sec:sup} and \ref{sec:gal}, 
respectively. Further discussion is made in \refsec{sec:con}.

\section{LTB models of super structures} \label{sec:ltb}

We model the structures as spherically symmetric, dust-filled
objects embedded in the EdS universe. The line element 
in such models is described by the LTB metric
\beq \label{eq:ltb}
ds^2 = -dt^2 + \frac{R'^2(t,r)dr^2}{1-K(r)r^2}+R^2(t,r)d\Omega^2,
\eeq
where $R(t,r)$ is the angular diameter distance of the coordinate 
$r$ as viewed from the center, $K(r)$ is the curvature 
function, a prime denotes a partial derivative with respect to $r$, 
and the speed of light has been set to unity. The evolution of 
$R(t,r)$ is determined by $K(r)$ and the mass function $M(r)$
\beq \label{eq:RdotS}
\dot{R}^2(t,r) = \frac{2GM(r)}{R(t,r)} - K(r)r^2,
\eeq
where $G$ is Newton's constant, and an overdot stands for a partial 
derivative with respect to $t$ (hereafter, we suppress the variables 
$t$ and $r$ if there is no ambiguity in the context). 
The mass function is related to the acceleration
\beq \label{eq:Rdd}
\ddot{R} = -\frac{GM}{R^2},
\eeq
and the coordinate density is given by
\beq \label{eq:rho}
\rho = \frac{M'}{4\pi R^2R'}.
\eeq

The general parametric solution of \refeq{eq:RdotS} for $K>0$ is
\eqa{ \label{eq:R}
R &= \frac{GM}{Kr^2}(1-\cos u) \\  \label{eq:t-tB}
t - t_B(r) &= \frac{GM}{K^{3/2}r^3}(u - \sin u).}
The bigbang time, $t_B(r)$, defines a coordinate surface, 
$R[t_B(r),r] = 0$, at bigbang and is determined by $M$ 
and $K$ up to a scaling of $r$. It ensures that every part 
of the inhomogeneous universe contracts to a singular point at 
the same coordinate time as we look back. The solution for 
$K < 0$ is obtained by replacing $u$ with $iu$.
In the special case, $K = 0$, we have 
\beq
R=\left(\frac{9GM}{2}\right)^{1/3}[t-t_B(r)]^{2/3}.
\eeq
If both $K$ and $M$ are constant, i.e., 
the universe is homogeneous and isotropic, 
the solution has the form $R = a r$ 
with $a$ being the scale factor of the cosmic expansion. 
More specifically, we reserve $a$ for the scale factor of the 
EdS universe, i.e., the background, in this paper.

A particular realization of the LTB model is generated by 
specifying $K$ (or $M$) and initial values of $R$ and $\dot{R}$.
Following \citet{marra07} and \citet{paranjape08}, we set 
\beq \label{eq:Ri}
R(t_i, r) = a_i r \quad\mbox{and}\quad \dot{R}(t_i, r) = a_i H_i r,
\eeq
where $a_i \equiv a(t_i)$ and $H_i \equiv H(t_i)$ are, 
respectively, the scale factor and Hubble expansion rate of the
EdS background at the initial time $t_i$. For convenience, we
choose $a_i = 10^{-3}$ and
$H_i a_i^{3/2} = H_0 = 50\,\mbox{km\,s}^{-1}\Mpc^{-1}$. 
This rather small value of the Hubble constant is often adopted in
the EdS universe to satisfy the age constraint, and, for this work,
it also roughly matches the angular size of the model structures 
in the EdS universe with that in \citealt{granett08} for the 
same linear size. 
Because we set the linear size of the structures in units of Mpc, 
a lower $H_0$ would make these structures smaller relative to 
the observable universe and reduce their RS signal. Results with
a larger $H_0$ are given in \refsec{sec:sup} for comparison.
\refeq{eq:Ri} ensures that the overdense (underdense) region always 
contracts (expands) against the background after $t_i$ (but before
the big crunch, if applicable) and thus 
becomes more and more overdense (underdense). One may also
set up voids with the overdensity in underdense region not 
always decreasing with time \citep*[e.g.,][]{alnes06a,kolb08}.

We study both compensated and uncompensated LTB models. 
The former requires the mean density of the region to match that
of the EdS background beyond a finite radius, and hence there is
no effect on the rest of the universe due to Birkhoff's theorem.
We design our toy models to have a smooth density profile
(i.e., at least continuous in its first derivative with respect
to $r$) with a uniform inner region (region I) inside a radius 
$\rI$, so that numerical results there can be 
checked against analytic solutions. 
The compensated model is realized by setting $K=0$ 
beyond an outer radius, $\rII$. We refer to the region between 
$\rI$ and $\rII$ as region II and that outside $\rII$ as region III,
which evolves exactly as the EdS universe.
For the uncompensated model, we set the initial coordinate density 
in region III to the initial EdS background density $\rhoEi$.

Because the geodesic equation involves $R''$, a discontinuous
$R''$ (such as those conforming with top-hat density profiles) 
will deflect photons in coordinate space (but not in physical
space) as they travel across the discontinuity. This does not 
have theoretical impact 
on the investigation, but to ensure the numerical precision, one 
must accurately determine the 4-coordinate $x^\mu$ where the 
photon crosses the discontinuity and then adjust its wavevector
$k^\mu$, so that its physical velocity remains the same and 
$k^\mu k_\mu = 0$ across the discontinuity. 
We avoid such a numerical issue by designing a 
curvature function that is continuous to at least second order
(so are $R$ and $M$), which renders smooth density profiles
via \refeq{eq:rho}. The functional forms of $K$ and corresponding
density profiles are given in \refapp{app:prf}. In summary, 
the parameters of the toy models are $\rI$, $\rII$,
and the initial overdensity of region I, 
$\dIi \equiv \rhoIi/\rhoEi - 1$.

Although there is no peculiarity analytically as the curvature 
function $K$ approaches 0, e.g., just inside of $\rII$ in the 
compensated models, Equations~(\ref{eq:R}) and (\ref{eq:t-tB}) are 
not suitable for numerical evaluation when $|K|$ is very small.
Observing that $q \equiv u |K|^{-1/2}$ remains finite as $|K|\to 0$, 
we Taylor expand Equations~(\ref{eq:R}) and (\ref{eq:t-tB}) around 
$u = 0$ and then replace $u$ with $q |K|^{1/2}$ when $K \sim 0$. 

\section{Geodesics} \label{sec:geo}

The geodesic equation is given by
\beq \label{eq:geo}
\frac{d^2x^\mu}{dv^2} + \Gamma_{\alpha\beta}^\mu 
\frac{dx^\alpha}{dv} \frac{dx^\beta}{dv} = 0,
\eeq
where $x^\mu=(t,r,\theta,\phi)$, $v$ is the affine parameter,
and $\Gamma_{\alpha\beta}^\mu$ is the Christoffel symbol. 
In the LTB metric, \refeq{eq:geo} becomes
\eqa{  \label{eq:tdd}
\frac{dk^t}{dv} = & -\frac{R'\dot{R}'}{1-Kr^2}\left(k^r\right)^2 
  -R\dot{R} \left(k^\Omega\right)^2 \\  
\frac{dk^r}{dv} = & -\frac{2\dot{R}'}{R'}k^tk^r
  -\left[\frac{R''}{R'}+\frac{2Kr+K'r^2}{2(1-Kr^2)}\right]
  \left(k^r\right)^2 \nonumber \\* \label{eq:rdd}
  & + R \frac{1-Kr^2}{R'} \left(k^\Omega\right)^2\\
\label{eq:hdd} \frac{dk^\theta}{dv} = &
  -\frac{2}{R}\frac{d R}{dv} k^\theta + 
  \sin\theta \cos\theta \left(k^\phi\right)^2 \\
\label{eq:pdd} \frac{dk^\phi}{dv} = &
  -\frac{2}{R}\frac{d R}{dv} k^\phi - 2\cot\theta k^\theta k^\phi,}
where $k^\mu \equiv d x^\mu/dv$ is the wavevector,
$\left(k^\Omega\right)^2 \equiv \left(k^\theta\right)^2 
  + \sin^2\theta \left(k^\phi\right)^2$, and
$dR/dv = \dot{R}k^t+R'k^r$.
We enumerate the index $\mu$ explicitly as $t$, $r$, $\theta$, and 
$\phi$ instead of numbers for clarity. 
Because of  spherical symmetry, we only need 
to study geodesics in the equatorial plane.

The frequency of a photon with wavevector $k^\mu$ as measured by 
an observer (or source) with 4-velocity $U^\mu$ is
\beq  \label{eq:nu}
\nu \propto -k^\mu U_\mu.
\eeq
If both the observer and the source are comoving in the LTB metric, 
then the photon redshift is given by
\beq \label{eq:z}
1 + z = \frac{k^t(t_\rme, r_\rme)}
{k^t(t_\rmo, r_\rmo)},
\eeq
where the subscripts e and o denote the event of emission and 
that of observation, respectively, and 
$k_\rme^t \equiv k^t(t_\rme, r_\rme) = 1$ is 
set arbitrarily as the initial condition. 
In actual calculations, we propagate photons backward from the 
observer to the source surface by setting $v\to -v$ (hence 
$k^\mu \to -k^\mu$) and the initial condition 
$k_\rmo^t \equiv k^t(t_\rmo, r_\rmo) = -1$. 
While Equations~(\ref{eq:geo}--\ref{eq:z}) remain the same under 
the reversal of $v$, others may change sign. 
To avoid confusion, we only refer to the forward case in all 
the equations and discussions below. 

The RS effect under investigation is of the order 
$|\Delta T|/T=|\Delta z|/(1+z) \sim 10^{-6}$--$10^{-5}$, so
one must ensure that numerical errors in $k^t$ is much less than 
one part in a million. Numerical results can be easily checked in 
uniform regions where alternative solutions exist. 
In addition, we perform several general tests that may
detect numerical errors, which are described as follows.

By definition, null geodesics obey $k_\mu k^\mu = 0$. This is 
a redundant constraint once the initial condition is set, as
in principle the geodesic equation does not induce violation of 
the condition. However, numerical errors could be accumulated.
While enforcing redundant constraints numerically is a
subject of research itself, we simply adjust 
the time steps so that $|k_\mu k^\mu / k_t k^t| \lesssim 10^{-8}$ 
when the photons exit the systems. This condition is not sufficient
to validate the results, but violation of it indicates significant
numerical errors.

In spherically symmetric systems, each geodesic remains in a plane.
One can always rotate the coordinates to place the geodesic in 
the equatorial plane, in which case $R^2 k^\phi$
is conserved. Similarly, $R^2 k^\theta$ is conserved if $k^\phi=0$.
We do not apply the known solution to reduce the dimensions of the 
system. Rather, we use them to check the precision of the numerical
solutions. The maximum fractional error of the conserved quantity 
along off-center geodesics is found to be $\sim 10^{-9}$.

Finally, radial geodesics can be calculated directly from the LTB 
metric
\refeq{eq:ltb}
\beq  \label{eq:dtdr}
\frac{dt}{dr} = \pm \sqrt{\frac{R'^2}{1-Kr^2}},
\eeq
where outward (inward) geodesics take the positive (negative) sign.
By differentiating \refeq{eq:z} with respect to $r(v)$, one gets 
the evolution of redshift  
\eqa{ \label{eq:dzdr} \nonumber 
\frac{d \ln (1+z)}{dr} &= -\frac{1}{k_\rmo^t}\frac{dk_\rmo^t}{dr}
  = -\frac{1}{k_\rmo^t} \left.\frac{dv}{dr}\right|_\rmo 
  \frac{d k_\rmo^t}{dv} \\
&= \frac{1}{k_\rmo^t k_\rmo^r}\Gamma_{\alpha\beta}^t 
  k_\rmo^\alpha k_\rmo^\beta = \pm 
  \frac{\dot{R}'}{\sqrt{1-Kr^2}}, }
where the sign on the far right side is the same as that in 
\refeq{eq:dtdr}
\citep[see][for an alternative derivation]{mustapha98,celerier00}.
Fractional differences between numerical solutions of $z(r)$ and 
$t(r)$ using Equations~(\ref{eq:tdd}--\ref{eq:rdd}) and those 
using Equations~(\ref{eq:dtdr}) and (\ref{eq:dzdr}) are
$\lesssim 10^{-8}$ when photons exit the model structures.

\section{Rees--Sciama effect of super structures} \label{sec:sup}

The clusters studied in \citealt{granett08} have radii of 
roughly $100\Mpch$ and a mean galaxy overdensity of 
0.72 within the overdense region, corresponding to a matter 
overdensity of 0.36 with a galaxy clustering bias of roughly
2 for luminous red galaxies \citep{padmanabhan07,blake08}. Thus,
our simple LTB models of these clusters have an inner radius
$\rI=140 \Mpc$\footnote{We adopt $h=0.71$ here only to obtain 
the size of the observed structures in units of Mpc.}, an outer 
radius $\rII = 280 \Mpc$ (205 Mpc) for the compensated 
(uncompensated) profile, and 
a uniform overdensity $\dI=0.36$ in region I at $z=0.5$. 
For voids, the only difference is $\dI=-0.19$.
\refapp{app:prf} provides more details about these profiles and their
corresponding metric quantity $R$.

To calculate the RS effect, we place a model structure between an 
observer and a source surface and propagate photons from the latter
to the former (as mentioned in \refsec{sec:geo}, they are actually
propagated backwards). The source surface is spherical around the 
observer in coordinate space, so that in absence of the structure, 
photons originated from the surface at the same time would reach the 
observer at the same time with the same redshift. With the 
intervening structure though, the photons received at the same time 
would have started from the source surface at different times 
because of gravitational time delay, and their redshifts would have
to compensate for the time delay plus additional changes because of 
nonlinear evolution of the structure. The net change 
in temperature is a sum of the two components:
\beq \label{eq:dT-t-z}
\Delta T = -T_\mathrm{CMB}\left[H_\mathrm{s} \Delta t_\mathrm{s} + 
\frac{\Delta z_\mathrm{s}}{1+z_\mathrm{s}}\right],
\eeq
where $T_\mathrm{CMB}=2.73\,\mathrm{K}$ is the present CMB temperature,
$\Delta t_\mathrm{s}$ is the time difference at the source,  
$H_\mathrm{s} \equiv H(z_\mathrm{s})$ is the Hubble parameter at the 
source redshift $z_\mathrm{s}$, and $\Delta z_\mathrm{s}$ is the redshift 
difference ``measured'' by an observer at present. Note that the two 
components in \refeq{eq:dT-t-z} cannot be measured separately in the 
case of the CMB. The time component accounts for the decrease of the 
CMB temperature with time, and $\Delta z_\mathrm{s}$ bears the 
subscript s because it varies with the source location. 

In the uniform background universe, a small time difference of 
$\Delta t$ translates to a redshift difference of
\[
\Delta z = - (1+z) H(z) \Delta t.
\]
Hence, in region III of compensated models, one can separate 
$z_\mathrm{s}$ into a time-compensation term 
\[
z_\mathrm{s}^{t} = -(1+z_\mathrm{s})H_\mathrm{s}\Delta t_\mathrm{s}
\] 
and an evolution term 
$z_\mathrm{s}^{e} = \Delta z_\mathrm{s} - z_\mathrm{s}^{t}$. 
Since $z_\mathrm{s}^{t}$ exactly cancels $H_\mathrm{s}\Delta t_\mathrm{s}$ 
in region III, \refeq{eq:dT-t-z} 
becomes
\beq \label{eq:dT-ze}
\Delta T = -T_\mathrm{CMB}\frac{\Delta z_\mathrm{s}^e}{1+z_\mathrm{s}}.
\eeq
In the compensated models, the only place where CMB photons are 
affected by the structure is within $\rII$. Hence, the RS effect
(and ISW effect in general) does not depend on the location of 
the hypothetical source surface as long as the compensated structure 
is enclosed between the observer and the source. In other words, 
$z_\mathrm{s}^e$ scales linearly with $1+z_\mathrm{s}$ in the 
uniform background. For the uncompensated 
models, the density evolution just beyond $\rII$ is still 
significantly different from that in the EdS universe, so we place 
the source far away from the model structure.

\begin{figure}
\centering
\epsscale{0.9}
\plotone{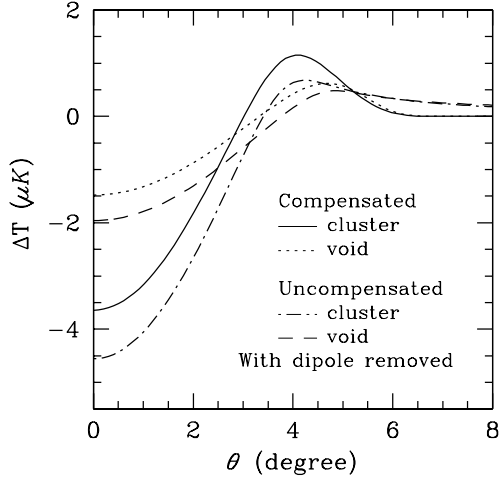}
\caption{CMB temperature shifts due to intervening model structures 
at $z = 0.5$, which have similar sizes ($\rI=140 \Mpc$ or
$\sim 4\deg$) and overdensities ($\dI = 0.36$, $-0.19$) as the 
clusters and voids in \citet{granett08}. The average densities 
of the compensated cluster (solid line) and void (dotted line) 
within the outer radius $\rII=280 \Mpc$ equal the background 
density. The uncompensated cluster (dot-dashed line) and void 
(dashed line) have very similar inner profiles ($r<1.4\rI$) as 
their compensated counterparts (see \reffig{fig:del-R}). The most 
prominent feature of the uncompensated models is a CMB dipole, 
which has been removed.
\label{fig:dtc}}
\end{figure}

\reffig{fig:dtc} illustrates the RS effect of four LTB model
structures. The main feature is a temperature decrement of several
$\muK$ in the inner region and a slight temperature increase at the 
outskirt, regardless whether the structure is a cluster or a
void. This is distinct from the linear ISW effect in a universe
dominated by dark energy, which generally causes a temperature increase 
for clusters and a decrement for voids. The RS effect of the voids may
seem counterintuitive at first, and we find that it is the result of 
underdensities generally evolving slower than the expansion rate in 
the EdS universe. 

Because a distant comoving observer or emitter in the uncompensated 
LTB models would be seen to
have a peculiar velocity in the EdS background, the most prominent 
feature of these models is a CMB temperature dipole. Once the dipole
is subtracted, the temperature profiles are similar to those of 
the compensated models.

As mentioned in \refsec{sec:ltb}, we set a low value for the Hubble
constant in the EdS background. If we used instead
$H_0=71\,\mbox{km\,s}^{-1}\Mpc^{-1}$, the central temperature decrement
of the compensated cluster (void) in \reffig{fig:dtc} would become
$9.9 \mu\mathrm{K}$ ($4.2 \mu\mathrm{K}$), nearly tripling the 
decrement of $3.6 \mu\mathrm{K}$ ($1.5 \mu\mathrm{K}$) with 
$H_0 = 50 \,\mbox{km\,s}^{-1}\Mpc^{-1}$.

\begin{figure}
\centering
\epsscale{0.9}
\plotone{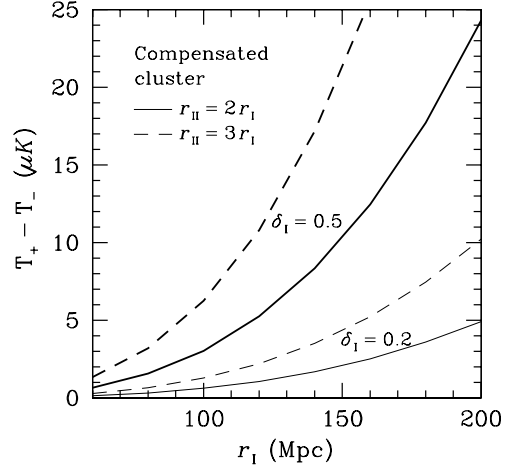}
\caption{Amplitude of the RS effect of compensated clusters as a function 
of the inner radius with $\rII = 2 \rI$ (solid lines) and $\rII = 3 \rI$
(dashed lines). Thick lines are for clusters with $\dI=0.5$ at 
$z=0.5$, while thin lines are for $\dI=0.2$.
\label{fig:dtm}}
\end{figure}

\reffig{fig:dtm} shows the dependence of the RS effect on the 
model parameters $\rI$, $\rII$, and $\dI$. We use the difference 
between the characteristic peak and trough of the temperature profile
($T_+ - T_-$) in \reffig{fig:dtc} 
to represent the amplitude of the RS effect and show 
results for compensated clusters. Because it takes more time 
for CMB photons to go through a larger structure, its potential can 
evolve more to produce stronger RS effects. Structures with larger
$|\dI|$ evolve faster and also produce larger RS effects. 
These are indeed seen in \reffig{fig:dtm}. Moreover, the RS effect 
increases fairly rapidly with $\rI$, $\rII$, and $\dI$.

The several $\muK$ level RS effect in \reffig{fig:dtc} is 
significant compared to the detected CMB temperature shift of 
$\sim \pm 10\muK$ due to super structures and is comparable 
to the expected maximum linear ISW effect of $4.2\muK$ 
within an aperture of $100\Mpch$ in the Millennium simulation
\citep{granett08}. Although the RS effect could bring the 
expected total ISW effect of the supervoids more in line 
with the observed value, it would also widen the discrepancy between 
the observation and the expected total ISW effect for 
superclusters. 

\begin{figure*}
\centering
\epsscale{0.85}
\plotone{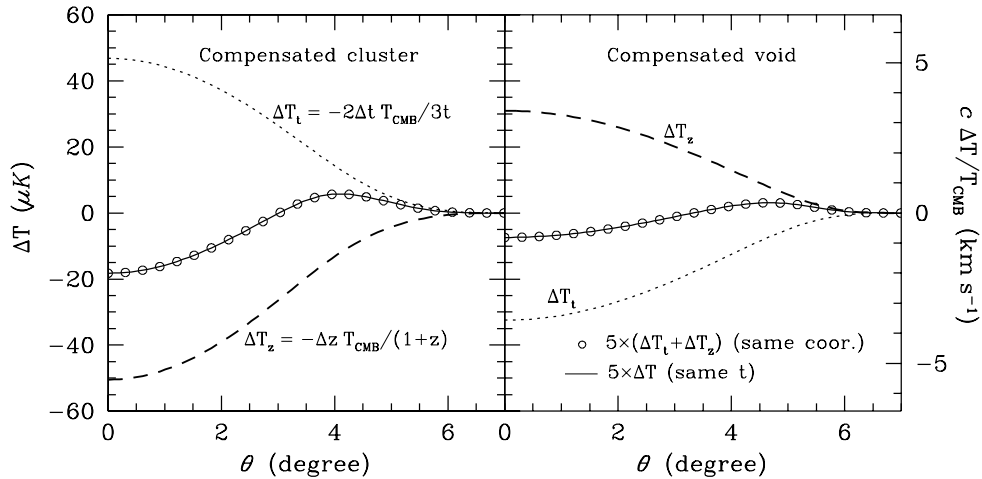}
\caption{Two equivalent views of the RS effect: a combination (circles)
of the redshift (dashed lines, $\Delta T_z$) and time delay (dotted 
line, $\Delta T_t$) effects on 
photons started at the same coordinate distance from the observer and 
a single redshift effect of photons started out at the same time (solid
lines), e.g., the instance of recombination. The left panel and the 
right panel are for the compensated cluster and the compensated void,
respectively. The combined RS results (circles) are the same as those 
in \reffig{fig:dtc} but amplified by 5 times for better viewing. 
For galaxies, their redshift changes are given by 
$(1+z) \Delta T_z / T_\mathrm{CMB}$, i.e., slightly more redshifted 
behind clusters and less behind voids in the EdS universe. 
\label{fig:rsalt}}
\end{figure*}

\section{Effect of evolving structures on galaxy redshifts} \label{sec:gal}

\reffig{fig:rsalt} illustrates the time delay component (dotted lines)
and the redshift component (dashed line) of the RS effect (circles) as
given in \refeq{eq:dT-t-z}. These two effects are opposite to each 
other and are of the same order, so that the net result is much smaller 
than either one. In other words, the gravitational time delay in the 
potential is a more dominant effect than the potential evolution for
super structures. 

Treating recombination as an instantaneous event, which suffices 
our purpose, one can view the RS effect and the ISW effect in
general as a single redshift effect without time delay. This view
is fully equivalent to that of \citet{rees68}. However, one is no
longer comparing photons that have started from the same spatial
coordinates at different times; rather, the CMB photons received 
at the same time were emitted at the same instance of recombination, 
but they must have started from different locations. 
In other words, the last scattering surface as viewed by the 
observer is no longer spherical in presence of inhomogeneities.

To cross-check the results, we trace photon geodesics 
to the same time surface rather than the same spatial coordinate 
surface described in \refsec{sec:sup}. This is analogous to theoretical
calculations of the ISW effect by integrating over (conformal) time.
The resulting change in the 
temperature is then given solely by the redshift difference, i.e., 
a change in $k^t$. As shown in \reffig{fig:rsalt}, this single 
redshift effect (solid lines) gives exactly the same RS effect as
that from the combination of the time delay and redshift effects 
(circles). 

Unlike CMB photons, light from a comoving galaxy must be emitted 
from the same coordinates whether there is an intervening structure 
or not. Furthermore, galaxy redshifts can be directly measured
through spectroscopy, so 
that the redshift effect is no longer mixed with the time delay 
effect. Without the large cancellation from time delay, changes of 
galaxy redshifts can be roughly an order of magnitude greater than 
fractional changes in the CMB temperature. For the clusters 
(voids) studied in \refsec{sec:sup}, the redshift (blueshift) effect
is merely several kilometers per second, or at the percent level of 
the large-scale rms bulk velocity, estimated to be a few hundred 
kilometers per second \citep[e.g.,][]{zhan02}. 
Nevertheless, it might be detectable 
statistically in the future and could be somewhat significant to 
large-scale redshift surveys that aim to precisely measure the
redshift distortion effect.

\section{Discussion and conclusions} \label{sec:con}

We have studied the intrinsic RS effect of super structures due to
their weakly nonlinear evolution using idealized LTB models and 
find that clusters and voids resembling those in 
\citet{granett08} can cause CMB temperature decrements of a few 
$\mu K$ in the EdS universe. The amplitude of the RS effect is
fairly significant compared to the measured CMB temperature 
changes, so one needs to take into account both the RS effect and 
the linear ISW effect when interpreting the observations. 

It is interesting to note that the RS effect is a central CMB
temperature decrement, regardless whether the structure is a 
cluster or a void. The reason is that generally overdensities grow 
faster than the cosmic expansion in the EdS universe, while 
underdensities evolves slower than the cosmic expansion.

In the concordant \lcdm{} universe, the linear ISW effect is 
expected to be stronger than the RS effect at low redshift when 
dark energy is dominant. A full calculation of 
the total ISW effect with \refeq{eq:geo} is rather involved, as 
the metric function $R$ and its 
first and second derivatives are no longer analytic in the \lcdm{}
universe. We can estimate the very effect along radial 
geodesics using Equations (\ref{eq:dtdr}) and (\ref{eq:dzdr}). The 
required $\dot{R}$ and $\dot{R}'$ evolve according to
\eqa{
\ddot{A} &= H_0^2 (1 - \Omega_\mathrm{m}) A - 
\frac{H_0^2 \Omega_\mathrm{m} + a_i K}{2 A^2} \\
\dot{A}' &= \frac{\ddot{A} A'}{\dot{A}} + 
\frac{K'}{2\dot{A}} \left(\frac{a_i}{A} - 1\right),
}
where $A = r^{-1}R$, and $\Omega_\mathrm{m} = 0.3$ is the present 
matter fraction of the background flat \lcdm{} universe. The initial 
conditions are given by \refeq{eq:Ri}, and the Hubble constant 
is set to $70\,\mbox{km\,s}^{-1}\Mpc^{-1}$ in this case. For the 
compensated cluster (void) in \reffig{fig:rsalt}, we find that 
$\Delta T_z = -21.5\,\mu\mathrm{K}$ ($13.3\,\mu\mathrm{K}$), 
$\Delta T_t = 26.6\,\mu\mathrm{K}$ ($-18.7\,\mu\mathrm{K}$), 
and the net temperature shift
$\Delta T = 5.1\,\mu\mathrm{K}$ ($-5.4\,\mu\mathrm{K}$) 
through its center. These results agree with the general expectation 
that photons gain (lose) energy going through a decaying potential 
well (wall) in the accelerating \lcdm{} universe.

For galaxies, the redshift effect is separable from the time delay 
effect, and the magnitude of the change in redshift is much larger
than the fractional change in the CMB temperature. This redshift 
effect is due in most part to the compensation for the time delay 
effect, e.g., a galaxy seen at an earlier time also has a higher
redshift. Evolution of the potential plays a minor role in the case
of super structures. Hence, the effect on the galaxy redshift is 
different from the RS effect or the ISW effect.

The compensated cluster (void) in \reffig{fig:rsalt} changes the 
redshift of a $z_\mathrm{s} \simeq 0.6$ background galaxy by 
$\Delta z_\mathrm{s} = 2.8\times10^{-5}$ ($-1.7\times10^{-5}$) in the 
EdS universe and by $\Delta z_\mathrm{s} = 1.2\times10^{-5}$ 
($-7.5\times10^{-6}$) 
in the \lcdm{} universe. Although measuring such redshift differences 
is not completely out of technical reach, the fact that there is not a 
standard redshift to compare with makes it impossible to measure this 
effect on individual galaxies. Since it is a systematic perturbation 
to galaxy redshifts, one might be able to detect the effect by 
comparing the galaxy redshift distribution with and without intervening
structures. However, many issues come into play. For example, 
magnification by the very foreground structures also affects how 
background galaxies are selected into the sample and thus change the 
galaxy redshift distribution perhaps more significantly.

While the absolute time delay due to foreground structures is not 
directly observable, one can obtain the relative time delay between 
images of a strongly lensed source either by actual measurements 
with a variable source or by estimation with a lens model. 
This could have allowed a short cut to the
\SLt{} that has to run for decades. Assuming for the 
moment that the potential does not evolve, i.e., $\Delta T=0$ in 
\refeq{eq:dT-t-z}, we have
\beq  \label{eq:dv}
\Delta v=\frac{c \Delta z_\mathrm{s}}{1+z_\mathrm{s}}
= -cH_\mathrm{s}\Delta t_\mathrm{s}
= -\frac{c \Delta t_0}{1+z_\mathrm{s}} H_\mathrm{s},
\eeq
where $\Delta t_0$ is the relative time delay observed at $z=0$.
For a source at $z_\mathrm{s} = 1$, the velocity difference 
corresponding to
a time delay of one year is $-1.8\,\mathrm{cm\,s}^{-1}$. 
The evolution of the lens potential also contributes to the 
redshift difference. 
Because lenses are more or less virialized structures, one might 
approximate them as static objects in physical space and determine
their potential evolution in the expanding universe. One might also 
stack CMB measurements behind a large number of 
similar lenses to determine the potential evolution. 

For comparison, the \SLt{} gives \citep{loeb98}
\beq  \label{eq:dv-sl}
\Delta v=-\frac{c \Delta t_0}{1+z_\mathrm{s}} 
\left[H_\mathrm{s}-H_0(1+z)\right].
\eeq
Here $\Delta t_0$ is the time between two epochs of measurements. 
In the \lcdm{} universe, $H_\mathrm{s} \lesssim H_0(1+z)$ at 
$z_\mathrm{s} \lesssim 2.5$, 
so the direction of the redshift change in the strong-lensing case
is opposite to the \SLt{} at $z_\mathrm{s} \lesssim 2.5$. The amplitude
in the strong-lensing case is almost an order of magnitude larger 
than that in the \SLt{}, making the former comparatively easier to 
measure for the same $\Delta t_0$. Moreover, the \SLt{} requires 
extremely high-precision absolute calibration of the instruments 
over a few decades, which is more difficult than high-precision 
relative measurements performed at the same time using the same 
instrument in the strong lensing case. 

Unfortunately, as \citet{loeb98} pointed out, differences in the
conditions of the images (e.g., shape, magnification, noise, etc.) 
of the same galaxy could already cause larger differences in the
redshift measurements; even more problematically, proper motion of 
the lens could induce a difference in image redshifts 
\citep{birkinshaw83} far exceeding that associated 
with the time delay. Therefore, even though the strong lensing case 
has a few advantages, it is still impractical to use the 
redshift difference and time delay between images of a strongly 
lensed source to measure the Hubble expansion at the source redshift.

\acknowledgements
This work was supported by the Bairen program from the Chinese 
Academy of Sciences, the National Basic Research Program of China 
grant No. 2010CB833000, and the National Natural Science Foundation 
of China grant No. 11033005.

\appendix

\section{Model Structures} \label{app:prf}
We require the curvature function $K$ of the model structures
to have a continuous second derivative with respect to $r$. 
For compensated models, we specify $K$ directly
\beq \label{eq:K}
K=\left\{ \begin{array}{ll} \KI & 0 \le r < \rI \\
\KI \left(\frac{\rII-r}{\rII-\rI}\right)^3\left[4 + 
3\frac{(r-2\rI+\rII)(2r-\rI-\rII)}{(\rII-\rI)^2}\right] & 
\rI \le r < \rII \\  0 & r \ge \rII \end{array} \right. ,
\eeq
where $\KI = \dIi {a_i^2 H_i^2} \Omega_\mathrm{m}$, and $\dIi$ is the 
initial overdensity of region I. For uncompensated models, we 
let the initial overdensity $\delta_i$ follow the same form 
as $K$ (with $\KI$ replaced by $\dIi$)
and then use Equations~(\ref{eq:rho}), 
(\ref{eq:Ri}), and (\ref{eq:RdotS}) to obtain $K$. We choose
$\rII$ for uncompensated models so that the uncompensated overdensity 
profiles roughly match those of compensated models in the central 
region. 

\begin{figure}
\centering
\epsscale{0.85}
\plottwo{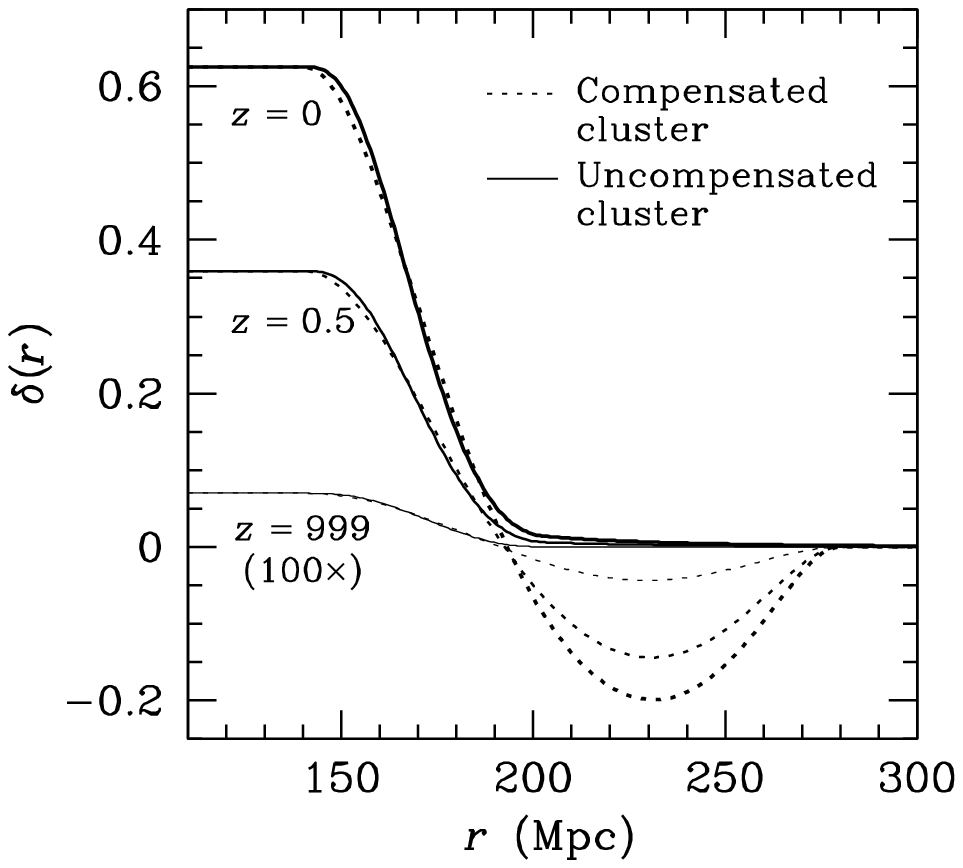}{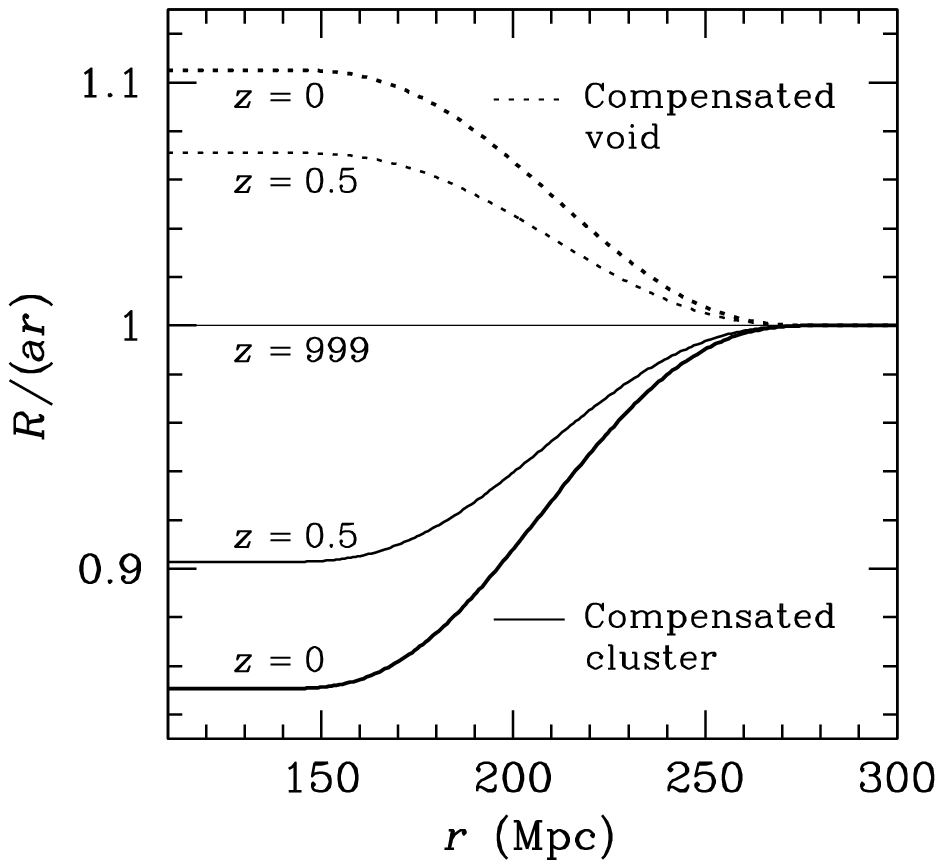}
\caption{\emph{Left panel}: 
Overdensity profiles of the compensated (dotted lines) and 
uncompensated (solid lines) cluster models. 
The compensated model has $\rI=140\Mpc$ and 
$\rII=280\Mpc$, while the uncompensated model has $\rI = 140\Mpc$
and $\rII=205\Mpc$. Profiles of the voids are similar to their 
corresponding cluster 
profiles but are inverted with lower amplitudes.
\emph{Right panel}: Evolution of the function $R/ar$ of the compensated 
cluster (solid lines) and the compensated void 
(dotted lines). The initial condition at 
$a_i = 10^{-3}$ sets $R_i=a_ir$. The uncompensated models 
behave similarly as their compensated counterpart
but converge to $R/ar = 1$ only at $r=\infty$.
\label{fig:del-R}}
\end{figure}

The left panel of \reffig{fig:del-R} shows overdensity profiles of the 
compensated (dotted lines) and uncompensated (solid lines) cluster 
models. For both models, 
the initial overdensity of region I at $a_i$ is set to 
$\dIi=7.0\times10^{-4}$, so that $\dI = 0.36$ at redshift $z=0.5$
to match the mean overdensity of overdense regions of the 
clusters found in \citet{granett08}. The density of the 
uncompensated cluster grows with time everywhere. The 
void models behave similarly (not shown) but with inverted 
profiles and smaller amplitudes. Moreover, the uncompensated 
void develops slightly positive overdensity 
($\delta \ll |\dI|$) at $r \gtrsim \rII$ at late time.

The right panel of \reffig{fig:del-R} illustrates the evolution of the 
function $R/ar$ for the compensated cluster (solid lines) and the 
compensated void (dotted lines). 
The void has $\dIi=-5.5\times 10^{-4}$ and 
$\dI = -0.19$ at $z = 0.5$, matching the mean 
overdensity of underdense regions of the voids in 
\citet{granett08}. In our setup, underdense regions always 
expand faster than the background after $t_i$. Hence,
for the voids, $R(t,r) \ge a r$ at $t>t_i$, and the equality
takes place where the mean overdensity within $r$ vanishes.
Similarly, for the clusters, $R(t,r) \le a r$ at $t >t_i$.


\end{document}